\documentclass{article}
\usepackage[preprint]{spconfa4}
\usepackage{amsmath,graphicx}

\usepackage[utf8x]{inputenc} 
\usepackage{amssymb}
\usepackage{color,soul}
\usepackage{hyperref}
\hypersetup{colorlinks=true}
\usepackage[textfont=it,tableposition=top]{caption}
\usepackage{multirow}
\usepackage[hang,flushmargin]{footmisc}
\usepackage{caption} 
\usepackage[textfont=it,tableposition=top]{caption}
\usepackage{hyperref,color,soul}
\hypersetup{colorlinks=true}

\newcommand\blfootnote[1]{%
  \begingroup
  \renewcommand\thefootnote{}\footnote{#1}%
  \addtocounter{footnote}{-1}%
  \endgroup
}

\copyrightnotice{978-1-6654-6867-1/22/\$31.00~\copyright2022 IEEE}
                   
\title{AUTHOR GUIDELINES FOR IWAENC 2022 PROCEEDINGS MANUSCRIPTS}
\title{STREAMING NOISE CONTEXT AWARE ENHANCEMENT FOR AUTOMATIC SPEECH RECOGNITION IN MULTI-TALKER ENVIRONMENTS}
\name{Joe Caroselli, Arun Narayanan, Yiteng Huang$^*$\thanks{$^*$Yiteng Huang is currently at Amazon but was with Google when this work was performed.}}
\address{Google LLC, U.S.A.}
\begin{document}
%
\maketitle
\begin{abstract}
One of the most challenging scenarios for smart speakers is multi-talker, when target speech from the desired speaker is mixed with interfering speech from one or more speakers. A smart assistant needs to determine which voice to recognize and which to ignore and it needs to do so in a streaming, low-latency manner. This work presents two multi-microphone speech enhancement algorithms targeted at this scenario. Targeting on-device use-cases, we assume that the algorithm has access to the signal before the hotword, which is referred to as the noise context. First is the Context Aware Beamformer which uses the noise context and detected hotword to determine how to target the desired speaker. The second is an adaptive noise cancellation algorithm called Speech Cleaner which trains a filter using the noise context. It is demonstrated that the two algorithms are complementary in the signal-to-noise ratio conditions under which they work well. We also propose an algorithm to select which one to use based on estimated SNR. When using 3 microphone channels, the final system achieves a relative word error rate reduction of 55\% at {-12~dB}, and 43\% at {12~dB}.
\end{abstract}


\blfootnote{    We thank Alex Gruenstein and Turaj Shabestary for several useful discussions and feedback and Jon Moeller and Brant Jameson for their help with data collection.}

\section{Introduction}
\label{sec:intro}

Automatic speech recognition (ASR) has shown robustness in the presence of noise largely due to the adoption of neural network based acoustic models \cite{PrabhavalkarRaoSainathLiEtAl17,BattenbergChenChildCoatesEtAl17,HoriWatanabeZhangChan2017,li2021betterfaster},large scale training  \cite{mirsamadi2017multi,hakkani2016multi,NarayananMisraSimPundakEtAl18}, and improved data augmentation strategies \cite{kim2017mtr, park2019specaugment, medennikov2018investigation}. Multiple speaker scenarios, however, still pose a challenge \cite{barker2017thirdchime,barker2018fifthchime}. This is particularly true for smart speakers where it is desired to respond to one of the speech sources and not the others.  In addition, smart speakers require streaming, low-latency solutions so the desired speaker needs to be quickly determined and isolated from the other sources. 

Solutions aimed at separating multiple speakers have been proposed using either a single  \cite{hershey2016deep,luo2019conv,luo2020dual} or multiple microphones \cite{chang2019mimo,chang2020end} are often designed for separating the multiple voices rather than identifying one target.  Filtering based on a targeted speaker has shown success but that targeted speaker needs to have registered with the device. \cite{wang2018voicefilter}  

A method which has shown considerable improvement isolates the desired source using a time-frequency mask and then uses the statistics of that masked source to steer a beamformer \cite{higuchi2016robust,heymann2016neural}. While effective in non-speech noise, it encounters challenges deciding between one or more voices. The entire utterance is usually used to steer the beamformer so the streaming, low-latency requirements of the smart speaker are not met. When operating under such constraints, these techniques have had difficulties  \cite{heymann2018performance,li2016neural}. 

Two techniques are studied to address speech-based noise while meeting streaming, low-latency criteria.  The device is assumed to trigger on a wakeword or hotword phrase such as ``OK Google" or ``Alexa" that precedes the user request or query. The desired speaker is, by definition, the one speaking the hotword. These techniques function on device, which enables them to make use of the signal that occurs directly before the hotword, referred to as the noise context.

The first is the Context Aware Beamformer (CAB) which uses the noise context and the detected hotword to determine the steering vector. The second is an adaptive noise cancellation algorithm called Speech Cleaner. It has resulted in significant improvement in hotword recognition in noisy environments\cite{huang2019hotword,huang2019multi}. Here it is applied to the query as well. Both will be compared to a beamformer steered with an ideal mask.

The rest of the paper is organized as follows. An overview of the system is given in Section~\ref{sec:model}. In Section~\ref{sec:cab}, we describe context aware beamforming. Section \ref{sec:cleaner} will describe the adaptive noise cancellation algorithm called the Speech Cleaner. Section \ref{sec:experiment} will describe the data used and the experimental setup. Evaluations will be provided in Section \ref{sec:results}.  
\section{Signal Model}
\label{sec:model}

Each received utterance consists of a hotword followed by the query as well as some audio before the hotword. The received short-time Fourier transform (STFT)-processed $M$ microphone signal is written as 
\begin{equation}
    \mathbf{Y}(k,n) \triangleq [Y_0(k,n), Y_1(k,n), ... Y_{(M-1)}(k,n) ]^T,
\end{equation}
where $Y_i(k,n)$ is the signal from microphone $i$ at frame $n$ for frequency subband $k$. The received signal can be written as the summation
\begin{equation}
    \mathbf{Y}(k,n) = \mathbf{X}(k,n) + \mathbf{V}(k,n),
\end{equation}
where $\mathbf{X}(k,n)$ is a vector of the M reverberant desired speech signals and $\mathbf{V}(k,n)$ corresponds to the noise signals which may also include interfering speech. Because all enhancement is conducted independently for each frequency subband, the frequency index $n$ shall subsequently be omitted.

\section{Context Aware Beamforming}
\label{sec:cab}
 A beamformer is a spatial filter applied to the $M$-microphone input signal to get an estimate of the desired signal
\begin{equation}
    {\hat{X}}(n) = \mathbf{W}^H\mathbf{{Y}}(n),
\end{equation}
where $\mathbf{W}$ is a vector of the $M$ coefficients of the beamformer and $\hat{X}(n)$ is a single channel estimate of the desired signal.

Perhaps the most challenging aspect of beamforming, especially in reverberant and multi-talker environments, is initializing the steering vector. using the principal eigenvector of the spatial correlation matrix of desired signal, which will collect the most energy of the desired signal, has been shown to be successful in reverberation\cite{souden2009optimal,van2004optimum}. In mask-based beamforming\cite{heymann2016neural, chang2019mimo}, neural networks are used to estimating this correlation matrix.  Multi-talker conditions, where there is one desired speaker and one or more interfering speakers, present challenges to ascertain which speaker it should be targeting. Doing so in a streaming, low-latency manner increases those challenges. 


The Context Aware Beamformer utilizes the hotword and the noise context to target the desired speaker. The spatial correlation matrix of the desired speaker $\mathbf{X}(n)$ is defined as:
\begin{equation}
\mathbf{\Phi_{XX}}=E_{n}[\mathbf{X}(n)\mathbf{X}(n)^H].
\end{equation}
This can be estimated by taking the difference between the correlation matrix calculated when there is both noise and signal and the correlation matrix when there is only noise:
\begin{equation} 
    \hat{\mathbf{\Phi}}_{\mathbf{XX}} \approx \mathbf{\Phi}_{\mathbf{YY}} - \mathbf{\Phi}_{\mathbf{VV}},
\label{eq:est}
\end{equation}
wherein it is assumed that the desired signal and the noise are uncorrelated \cite{taherian2019deep}.

In order to make use of this approximation, time periods that are only noise and that are signal plus noise must be identified. To that end, this work makes uses of the hotword. When the device identifies the hotword, it is, by definition, the desired speaker saying the hotword. Because noise will likely also be present, the period of time that is detected as containing the hotword can be used to estimate $\mathbf{\Phi}_\mathbf{YY}$.

Targeting on-device application, we assume that the model has access to the noise context, the period of time before the hotword. This segment of time can be assumed to contain only noise and that noise will have similar statistics to the noise in the hotword.  This can be used to estimate
$\mathbf{\Phi_{\mathbf{VV}}}$.

With these two estimates, Equation (\ref{eq:est}) can be applied to obtain an estimate of the spatial correlation of the desired signal. The principal eigenvector of this matrix is then used as the initial beamformer coefficients:
\begin{equation}
    \mathbf{\hat{W}}=Pr\{\hat{\mathbf{\Phi}}_{\mathbf{XX}} \},
\end{equation}
where, $Pr\{\centerdot\}$ signifies finding the principal eigenvector of the matrix. The beamformer coefficients can then be adapted using least mean squares (LMS) in accordance with the minimum variance distortionless response (MVDR) criterion \cite{frost1972algorithm}.

\section{Speech Cleaner}
\label{sec:cleaner}

Speech Cleaner (SC), an adaptive noise cancellation algorithm, is summarized in this section\cite{huang2019hotword, huang2019multi}. In  previous works, it has been shown to result a significant increased in robustness to noise in hotword detection. Here, it is to applied it directly to the target query in addition to the hotword.

A finite impulse response (FIR) filter with a tapped delay line of length $L$ is applied to the signals from all microphones except for one. The summed output of these is subtracted from the received signal at the first microphone:
\begin{equation}
    Z_m(n)=Y_0(n)-\sum_{m=1}^{M-1}{\mathbf{U}_m}^H\tilde{\mathbf{Y}}_m(k,n),
\end{equation}
where,
\begin{equation}
\tilde{\mathbf{Y}}_m(n) = [Y_m(n), Y_m(n-1), \cdots Y_m(n-(L-1))]^T,
\end{equation}
is a vector of time delayed STFT-processed input for microphone $m$ and,
\begin{equation}
    \mathbf{U}_m(k) = [U_m(k,0), U_m(k,1), \cdots U_m(k,L-1) ]^T 
\end{equation}
is a vector of the filter coefficients to be applied to microphone input $m$. The filter coefficients are specified as those that minimize the power of the enhanced output
\begin{equation}
    \hat{\mathbf{U}}_m=\arg \underset{\mathbf{U}_m}{\min}  E_n{[|Z_m(n)|^2]}.
\end{equation}

The minimization is done through adaptation during the noise context, when it is assumed there is no desired speech present. In this way, the filter becomes trained to cancel out the noise sources. When the hotword is detected, the filter coefficients are frozen. The last coefficients before the hotword detection are then applied to the query:
\begin{equation}
    \hat{{X}}(n)=Y_0(n)-\sum_{m=1}^{M-1}{\mathbf{\hat{U}}_m}^H\tilde{\mathbf{Y}}_m(k,n-l)
\end{equation}
to obtain a single channel estimate of the desired signal. In this way, the filter works to cancel the noise, as it was trained, and not the desired signal, which was not present during training. Unlike the beamformer, Speech Cleaner has no constraint as to the amount of speech distortion introduced. 
\section{Experimental Setup}
\label{sec:experiment}

This study uses data re-rerecorded in a living-room lab with no sound treatment. Desired speech and noise were recorded separately with a three microphone array with two microphones on the top spaced 7.1~cm apart with a third was on the front. The queries were played r from 7 different speaker positions at a height of 1.5m and at 4m from the array. From each location 100 different queries prefaced with the hotword ``OK Google" or ``Hey Google" were played at a volume of 40 dB over ambient room noise. Two types of noise, speech-based from a movie and non-speech pink noise, were separately played through a loudspeaker from the same 7 locations. 

Speech and noise were mixed at several different SNR values ranging from -12 to 24 dB as well as a set with no added noise.  In all cases, the recordings contain whatever ambient noise was in the room during recording. They were mixed such that the speech and noise did not originate from the same speaker position and with 8s of noise recorded before the hotword commenced. The multi-channel input was processed by the specified enhancement algorithm to produce one channel of output which was then passed onto an ASR model. 

A state-of-the-art end-to-end recurrent neural transducer model is used for evaluations\cite{sainath2020streaming}. The model was trained using anonymized, hand-transcribed English utterances from domains like VoiceSearch, Farfield, Telephony and YouTube. Data augmentation was applied via a room simulator\cite{kim2017mtr} to model reverberation times from 0 ms and 900 ms and SNRs from 0 dB to 30 dB. Consequently, this model is relatively robust to moderate noise conditions.
\section{Results}
\label{sec:results}

\begin{table}[ht]
\vspace{-0.1in}
\caption{Two Channel Results} 
\vspace{-0.1in}
\centering 
\begin{tabular}{c | c c c c | c} 
\hline\hline 
SNR & Baseline & CAB & SC & SNR & Oracle \\
(dB) & &  & & Sel. & Bmfm\\ [0.5ex] 
\hline 
-12 & 100.2 & 95.4 & \textbf{64.5} & 68.0 & 95.7 \\ 
-6 & 83.0 & 56.6 & \textbf{43.6} & 48.8 & 66.0\\
0 & 48.7 & 34.4 & \textbf{24.7} & 28.3  & 29.7\\
6 & 20.5 & 17.9 & \textbf{12.0} & 12.4 & 11.7\\
12 & 9.2 & 6.8 & 7.6 & \textbf{5.6} & 5.3\\
$Clean$ & 3.0 & \textbf{1.9} & 3.0 & \textbf{1.9} & 1.8\\ [1ex] 
\hline 
\end{tabular}
\label{table:twoch} 
\vspace{-0.1in}
\end{table}

\begin{table}[ht]
\caption{Three Channel Results} 
\vspace{-0.1in}
\centering 
\begin{tabular}{c | c c c c | c } 

\hline\hline 
SNR & Baseline & CAB & SC & SNR & Oracle\\
(dB) & &  & & Sel. & Bmfm\\ [0.5ex] 
\hline 
-12 & 100.2 &  93.0 & \textbf{42.0} & 45.4 & 92.6\\ 
-6 & 83.0 & 66.2 & \textbf{28.9} & 33.0 & 60.4\\
0 & 48.7 & 31.7 & \textbf{16.6} & 20.1 & 25.3\\
6 & 20.5 & 11.6 & \textbf{10.9} & 11.1 & 10.1\\
12 & 9.2 & \textbf{4.6} & 8.0 & 5.2 & 4.3 \\
$Clean$ & 3.0 & \textbf{1.4} & 3.9 & \textbf{1.4} & 1.6\\ [1ex] 
\hline 
\end{tabular}
\vspace{-0.1in}
\label{table:threech} 
\vspace{-0.1in}
\end{table}

\vspace{-0.1in}
\subsection{Context Aware Beamformer}
\vspace{-0.1in}

CAB was tested with multi-talker noise datasets using two and three microphone configurations. No adaptation was done on the beamformer coefficients after the initialization with principal eigenvector. The word error rate (WER) results are presented for the two channel case in the third column of Table \ref{table:twoch} and the three channel case in Table \ref{table:threech}. An SNR value of $Clean$ indicates the case where there was no added multi-talker noise. Also added for comparison is the baseline case where the ASR model was applied directly with no enhancement.  In addition, there is an Oracle case where the spatial correlation matrix of the desired signal is calculated directly from the isolated desired signal.  This is the ideal case for mask-based beamforming.  

It can be seen that at the lowest SNR values the beamformer provided little benefit. In these cases, where the noise power is much larger than the signal, it can be difficult to get an accurate approximation for the desired signal correlation from \ref{eq:est}. In contrast, at 6 dB and above the relative WER improvements approach 50\% where it approaches the performance of the Oracle. Going from two to three microphones provides a notable improvement. 

\vspace{-0.1in}
\subsection{Speech Cleaner}
\vspace{-0.1in}
The fourth column of Tables \ref{table:twoch} and \ref{table:threech} show the performance of Speech Cleaner with the multi-talker data sets using two and three microphones, respectively. The filter length was $L=3$ with adaption using recursive least squares (RLS) \cite{huang2019multi} during the 8s noise context. The coefficient values at the detected hotword start time are then used to filter the hotword and query. 

The Speech Cleaner gives the most improvement at the lower SNR values where it significantly outperforms the Oracle beamformer. At -12dB, there is a 40\% relative WER improvement with two microphones, improving to 60\% with a third mic. However, at the higher SNR values, improvement is lower and  pales in comparison to the beamformer. In the no added noise case, adding an additional third microphone actually causes a performance degradation. Speech Cleaner, unlike CAB, does not contain any constraints with regard to signal distortion. In lower noise, the noise cancellation benefits are outweighed by the signal distortion. 

\vspace{-0.1in}
\subsection{SNR Based Selection}
\vspace{-0.1in}

The third and fourth columns in Tables \ref{table:twoch} and \ref{table:threech} show the CAB and Speech Cleaner to be complementary in the SNR ranges at which they best perform. Therefore, it is desirable to select the appropriate algorithm in the corresponding conditions. 

A simple method to estimate the SNR of the utterance is used. That estimate is used to select between the two enhancement techniques. The SNR estimate is obtained by separately measuring the power in the hotword segment and the power in the noise context. Because the hotword contains both signal and noise, its power is an estimate of the power of the signal plus that of the noise. The context power is an estimate of the noise power. Consequently,
\begin{equation}
\widehat{SNR}=\frac{\sigma_{hotword}^2}{\sigma_{context}^2}-1=\frac{\hat{\sigma}_X^2+\hat{\sigma}_V^2}{\hat{\sigma}_V^2}-1.
\end{equation}
If $\widehat{SNR}$ is less than a threshold $\gamma$ then the Speech Cleaner will be used otherwise CAB will be applied.

The fifth column of Tables \ref{table:twoch} and \ref{table:threech} shows results for the two microphone array with $\gamma$ at 6 dB. With the SNR-based selector, performance approaches that of the Speech Cleaner for SNR values of 6~dB and below. Above 6~dB, performance approaches that of CAB. This yields better than a 30\% and 43\%WER improvement across the considered SNR range for two and three channels, respectively.

\vspace{-0.1in}
\subsection{Non-Speech Noise}
\vspace{-0.1in}

To consider performance in non-speech noise, Table \ref{table:pink} presents WER for pink noise with three channels. The SNR-based selection tracks the performance of the Speech Cleaner past the 6dB threshold leading to some undesired results. This suggests that either the the thresholds could benefit from more tuning, or alternative methods of selection between the techniques should be considered. We leave that to future work.
\begin{table}[b]
\vspace{-0.1in}
\caption{Three Channel Pink Noise Results} 
\vspace{-0.1in}
\centering 
\begin{tabular}{c | c c c c  } 
\hline\hline 
SNR & Baseline & CAB & SC & SNR \\
(dB) & &  &  & Sel.\\ [0.5ex] 
\hline 
-12 & 100.2 & 100.0 & \textbf{42.2} & \textbf{42.2}\\ 
-6 & 98.9 & 89.3 & \textbf{25.5} & \textbf{25.5} \\
0 & 80.1 & 53.0 & \textbf{18.0} & \textbf{18.0} \\
6 & 42.2 & 18.0 & \textbf{12.3} & \textbf{12.3} \\
12 & 12.4 & \textbf{5.3} & 8.8 & 9.1 \\
18 & 3.3 & \textbf{2.5} & 8.6 & 5.6 \\
24 & 1.7 & \textbf{1.5} & 8.3 & 1.6 \\[1ex] 
\hline 
\vspace{-0.1in}
\end{tabular}
\vspace{-0.1in}
\label{table:pink} 
\end{table}
\vspace{-0.1in}

\subsection{Impact of Context Length}
\vspace{-0.1in}

The results thus far used all 8s of noise context. In Table \ref{table:context}, WERs are shown across the SNR range for shorter noise context lengths. It can be seen that, particularly at the higher SNRs where CAB is more effective, reducing the noise context to 1s still captures the bulk of the improvements. Benefits are still seen if only 0.25s of context is available. 

\begin{table}[t]
\vspace{-0.1in}
\caption{Impact of Noise Context Length on CAB} 
\vspace{-0.1in}
\centering 
\begin{tabular}{c | c c c c c c } 
\hline\hline 
Length & -12dB & 6dB& 0dB & 6dB & 12dB & $\infty$  \\ [0.5ex] 
\hline 
Baseline & 101.6 & 81.6 & 46.2 & 19.3 & 7.7 & 1.9 \\  
All(8s) & 95.7 & 71.5 & 33.9 & 12.7 & 5.6 & 1.9 \\ 
3s & 95.7 & 71.8 & 34.3 & 13.7 & 5.9 & 1.9 \\
1s & 97.3 & 74.5 & 36.6 & 14.0 & 6.0 & 1.9 \\
0.5s & 98.0 & 75.3 & 37.6 & 14.0 & 6.0 & 1.9 \\
0.25s & 98.6 & 76.5 & 39.1 & 14.9 & 6.2 & 2.1 \\ [1ex]
\hline 
\end{tabular}
\vspace{-0.1in}
\label{table:context} 
\vspace{-0.1in}
\end{table}

\subsection{Impact of Desired Speaker in Context}
\vspace{-0.1in}

It has been assumed that the desired speaker is not present during the noise context. While that is valid often, it will not always be the case. Here, the algorithms operate with the desired speaker talking throughout the context period. To simulate this, part of the query is appended before the hotword. In Table \ref{table:dic} the impact on the performance of CAB and Speech Cleaner for these cases with the two channel array is shown.

These results demonstrate the importance of ensuring that this assumption holds or allowing for the disabling of these algorithms when it does not. 

\begin{table}[ht]
\vspace{-0.1in}
\caption{Impact of Desired Speaker in Context} 
\vspace{-0.1in}
\centering 
\begin{tabular}{c c c c c } 
\hline\hline 
SNR & Baseline & CAB & Speech  \\
(dB) & &  &Cleaner & \\ [0.5ex] 
\hline 
-12 & 101.6& 97.9 & 77.8\\ 
-6 & 81.6 & 82.0 & 69.9 \\
0 & 46.2& 53.3 & 66.1 \\
6 & 19.3 & 28.8 & 67.8\\
12 & 7.7& 15.5 & 75.2\\
$Clean$ & 1.9& 4.6 & 91.0  \\ [1ex] 
\hline 
\end{tabular}
\vspace{-0.1in}
\label{table:dic} 
\end{table}
\vspace{-0.1in}

\section{Conclusion and Future Work}
\label{sec:conclusion}
Two techniques were presented to help address the multi-talker scenario on smart speakers. The Context Aware Beamformer used the noise context and detected hotword to determine how to target the desired speaker. CAB was shown to enable WER reduction across the SNR range, but it was most effective at higher SNRs. We also present an adaptive noise cancellation algorithm called Speech Cleaner trained using the noise context. This algorithm was shown to be very effective at SNRs below 6 dB where it outperformed an oraacle mask-based beamformer, but could degrade performance at higher SNRs. It was demonstrated that the two algorithms are complementary in the conditions under which they work well. To that end, an algorithm to select which algorithm to use based on estimated SNR estimation was demonstrated. With the SNR-based selection, relative error rates reductions of more than 30\% were demonstrated for a two microphone array, improving to more than 43\% for three microphones. 

Future work will encompass more advanced techniques to select between the two algorithms and the original raw audio. Identifying conditions when the target speaker also appears in the noise context is another interesting direction for the future.
\ninept
\bibliographystyle{IEEEbib}
\bibliography{refs}

\begin{thebibliography}{10}

\bibitem{PrabhavalkarRaoSainathLiEtAl17}
R.~Prabhavalkar, K.~Rao, T.~N. Sainath, B.~Li, L.~Johnson, and N.~Jaitly,
\newblock ``{A Comparison of Sequence-to-Sequence Models for Speech
  Recognition},''
\newblock in {\em Proc. Interspeech}, 2017.

\bibitem{BattenbergChenChildCoatesEtAl17}
E.~{Battenberg}, J.~{Chen}, R.~{Child}, A.~{Coates}, Y.~G.~Y. {Li}, H.~{Liu},
  S.~{Satheesh}, A.~{Sriram}, and Z.~{Zhu},
\newblock ``{Exploring Neural Transducers for End-to-end Speech Recognition},''
\newblock in {\em Proc. ASRU}, 2017.

\bibitem{HoriWatanabeZhangChan2017}
T.~Hori, S.~Watanabe, Y.~Zhang, and W.~Chan,
\newblock ``{Advances in Joint CTC-Attention Based End-to-End Speech
  Recognition with a Deep CNN Encoder and RNN-LM},''
\newblock in {\em Proc. Interspeech}, 2017.

\bibitem{li2021betterfaster}
B.~Li, A.~Gulati, J.~Yu, T.~N. Sainath, C.-. Chiu, A.~Narayanan, S.-Y. Chang,
  et~al.,
\newblock ``A better and faster end-to-end model for streaming asr,''
\newblock in {\em Proc. ICASSP}, 2021.

\bibitem{mirsamadi2017multi}
S.~Mirsamadi and J.~HL Hansen,
\newblock ``On multi-domain training and adaptation of end-to-end rnn acoustic
  models for distant speech recognition,''
\newblock in {\em Proc. Interspeech}, 2017.

\bibitem{hakkani2016multi}
D.~Hakkani-T{\"u}r, G.~T{\"u}r, A.~Celikyilmaz, Y-N. Chen, J.~Gao, L.~Deng, and
  Y.-Y. Wang,
\newblock ``Multi-domain joint semantic frame parsing using bi-directional
  rnn-lstm.,''
\newblock in {\em Proc. Interspeech}, 2016.

\bibitem{NarayananMisraSimPundakEtAl18}
A.~{Narayanan}, A.~{Misra}, K.~C. {Sim}, G.~{Pundak}, A.~{Tripathi},
  M.~{Elfeky}, P.~{Haghani}, T.~{Strohman}, and M.~{Bacchiani},
\newblock ``{Toward Domain-Invariant Speech Recognition via Large Scale
  Training},''
\newblock in {\em Proc. of SLT}, 2018.

\bibitem{kim2017mtr}
C.~Kim, A.~Misra, K.~Chin, T.~Hughes, A.~Narayanan, T.~Sainath, and
  M.~Bacchiani,
\newblock ``Generation of large-scale simulated utterances in virtual rooms to
  train deep-neural networks for far-field speech recognition in {Google
  Home},''
\newblock in {\em Proc. Interspeech}, 2017.

\bibitem{park2019specaugment}
D.~S. Park, W.~Chan, Y.~Zhang, C.-C. Chiu, B.~Zoph, E.~D. Cubuk, and Q.~V. Le,
\newblock ``Specaugment: A simple data augmentation method for automatic speech
  recognition,''
\newblock {\em Proc. Interspeech}, 2019.

\bibitem{medennikov2018investigation}
I.~Medennikov, Y.~Y. Khokhlov, A.~Romanenko, D.~Popov, N.~A. Tomashenko,
  I.~Sorokin, and A.~Zatvornitskiy,
\newblock ``An investigation of mixup training strategies for acoustic models
  in {ASR},''
\newblock in {\em Proc. Interspeech}, 2018.

\bibitem{barker2017thirdchime}
J.~Barker, R.~Marxer, E.~Vincent, and S.~Watanabe,
\newblock ``The third {CHiME} speech separation and recognition challenge:
  Analysis and outcomes,''
\newblock {\em Computer Speech \& Language}, vol. 46, pp. 605--626, 2017.

\bibitem{barker2018fifthchime}
J.~Barker, S.~Watanabe, E.~Vincent, and J.~Trmal,
\newblock ``The fifth'chime'speech separation and recognition challenge:
  Dataset, task and baselines,''
\newblock in {\em Proc. Interspeech}, 2018.

\bibitem{hershey2016deep}
J.~R. Hershey, Z.~Chen, J.~Le Roux, and S.~Watanabe,
\newblock ``Deep clustering: Discriminative embeddings for segmentation and
  separation,''
\newblock in {\em Proc. ICASSP}. IEEE, 2016, pp. 31--35.

\bibitem{luo2019conv}
Y.~Luo and N.~Mesgarani,
\newblock ``Conv-tasnet: Surpassing ideal time--frequency magnitude masking for
  speech separation,''
\newblock {\em IEEE/ACM transactions on audio, speech, and language
  processing}, vol. 27, no. 8, pp. 1256--1266, 2019.

\bibitem{luo2020dual}
Y.~Luo, Z.~Chen, and T.~Yoshioka,
\newblock ``Dual-path rnn: efficient long sequence modeling for time-domain
  single-channel speech separation,''
\newblock in {\em Proc. ICASSP}. IEEE, 2020, pp. 46--50.

\bibitem{chang2019mimo}
X.~Chang, W.~Zhang, Y.~Qian, J.~Le Roux, and S.~Watanabe,
\newblock ``Mimo-speech: End-to-end multi-channel multi-speaker speech
  recognition,''
\newblock in {\em Proc. ASRU}. IEEE, 2019, pp. 237--244.

\bibitem{chang2020end}
X.~Chang, W.~Zhang, Y.~Qian, J.~Le Roux, and S.~Watanabe,
\newblock ``End-to-end multi-speaker speech recognition with transformer,''
\newblock in {\em Proc. ICASSP}. IEEE, 2020, pp. 6134--6138.

\bibitem{wang2018voicefilter}
Q.~Wang, H.~Muckenhirn, K.~Wilson, P.~Sridhar, Z.~Wu, J.~Hershey, R.~A.
  Saurous, R.~J .Weiss, Y.~Jia, and I.~L. Moreno,
\newblock ``Voicefilter: Targeted voice separation by speaker-conditioned
  spectrogram masking,''
\newblock {\em arXiv preprint arXiv:1810.04826}, 2018.

\bibitem{higuchi2016robust}
T.~Higuchi, N.~Ito, T.~Yoshioka, and T.~Nakatani,
\newblock ``Robust mvdr beamforming using time-frequency masks for
  online/offline asr in noise,''
\newblock in {\em Proc. ICASSP}. IEEE, 2016, pp. 5210--5214.

\bibitem{heymann2016neural}
J.~Heymann, L.~Drude, and R.~Haeb-Umbach,
\newblock ``Neural network based spectral mask estimation for acoustic
  beamforming,''
\newblock in {\em Proc. ICASSP}. IEEE, 2016, pp. 196--200.

\bibitem{heymann2018performance}
J.~Heymann, M.~Bacchiani, and T.~N. Sainath,
\newblock ``Performance of mask based statistical beamforming in a smart home
  scenario,''
\newblock in {\em Proc. ICASSP}. IEEE, 2018, pp. 6722--6726.

\bibitem{li2016neural}
B.~Li, T.N. Sainath, R.~J. Weiss, K.~W. Wilson, and M.~Bacchiani,
\newblock ``Neural network adaptive beamforming for robust multichannel speech
  recognition,''
\newblock 2016.

\bibitem{huang2019hotword}
Y.~Huang, T.~Z. Shabestary, and A.~Gruenstein,
\newblock ``Hotword cleaner: dual-microphone adaptive noise cancellation with
  deferred filter coefficients for robust keyword spotting,''
\newblock in {\em Proc. ICASSP}. IEEE, 2019, pp. 6346--6350.

\bibitem{huang2019multi}
Y.~Huang, T.~Z. Shabestary, A.~Gruenstein, and L.~Wan,
\newblock ``{Multi-Microphone Adaptive Noise Cancellation for Robust Hotword
  Detection},''
\newblock in {\em Proc. Interspeech 2019}, 2019, pp. 1233--1237.

\bibitem{souden2009optimal}
M.~Souden, J.~Benesty, and S.~Affes,
\newblock ``On optimal frequency-domain multichannel linear filtering for noise
  reduction,''
\newblock {\em IEEE Transactions on audio, speech, and language processing},
  vol. 18, no. 2, pp. 260--276, 2009.

\bibitem{van2004optimum}
H.L. Van~Trees,
\newblock {\em Optimum Array Processing: Part IV of Detection, Estimation, and
  Modulation Theory},
\newblock Detection, Estimation, and Modulation Theory. Wiley, 2004.

\bibitem{taherian2019deep}
H.~Taherian, Z-Q Wang, and D.~Wang,
\newblock ``Deep learning based multi-channel speaker recognition in noisy and
  reverberant environments,''
\newblock in {\em Interspeech}, 2019.

\bibitem{frost1972algorithm}
O.~L. Frost,
\newblock ``An algorithm for linearly constrained adaptive array processing,''
\newblock {\em Proceedings of the IEEE}, vol. 60, no. 8, pp. 926--935, 1972.

\bibitem{sainath2020streaming}
T.~N. Sainath, Y.~He, B.~Li, A.~Narayanan, R.~Pang, A.~Bruguier, S.~Chang,
  W.~Li, R.~Alvarez, Z.~Chen, et~al.,
\newblock ``A streaming on-device end-to-end model surpassing server-side
  conventional model quality and latency,''
\newblock in {\em Proc. ICASSP}. IEEE, 2020, pp. 6059--6063.

\end{thebibliography}

\end{document}